\documentclass[final,3p,times,twocolumn]{elsarticle}

\usepackage{amssymb}
\usepackage{amsmath}

\journal{Physics Letters A}

\begin{document}

\begin{frontmatter}



\title{Polylogarithmic Structure of Bragg Diffraction in Finite-Coherence Lattices}


\author{Evangelos G. Filothodoros} 

\affiliation{organization={Physics Department, Aristotle University of Thessaloniki, Thessaloniki, Greece}}

\begin{abstract}
We develop a polylogarithmic structure for Bragg diffraction based on a weighted multi-plane interference model. Within this kind of construction, the scattering amplitude is expressed as a polylogarithmic generating function. By introducing extra contributions with power-law and the usual exponential decay, it takes the form $F(\theta) = \mathrm{Li}_m\left(e^{i\theta_{\mathrm{eff}} - \epsilon}\right)$, where $\epsilon$ is a finite coherence length. In the limit where $\epsilon \rightarrow 0$, the argument of the polylogarithm approaches the unit circle and the classical Bragg condition corresponds to the approach of the polylogarithm argument toward its branch point $z=1$. This formulation provides a compact analytical framework for describing diffraction line shapes within a generalized correlation model in which peak positions, widths, and line shapes arise from a single analytic structure. Although we are able to recover the standard Bragg law for ideal crystals, the polylogarithm model captures deviations due to finite correlation length, disorder and non-uniform lattice coherence. We show that if Bragg peaks correspond to boundary singularities of the polylogarithm, a connection between diffraction theory and complex analysis arise. The proposed theoretical model may be particularly relevant for disordered or partially coherent materials, where conventional diffraction models often require additional phenomenological broadening assumptions.
\end{abstract}

\begin{graphicalabstract}
\end{graphicalabstract}

\begin{highlights}
\item Research highlight 1
\item Research highlight 2
\end{highlights}


\end{frontmatter}



\section{Introduction}

X-ray diffraction remains one of the most fundamental tools for probing crystalline structure, with the Bragg law \cite{Egami}
\begin{equation}
n\lambda = 2d \sin\theta
\label{original}
\end{equation}
providing a remarkably accurate description of diffraction peak positions in periodic systems. This relation follows from the constructive interference of waves scattered by equally spaced lattice planes and has been extensively validated across a wide range of crystalline materials. However, the classical formulation assumes perfect periodicity, infinite coherence, and uniform scattering contributions, conditions that as we know are not always realized in experimental materials.

In practice, diffraction patterns exhibit finite peak widths, asymmetries and, in some cases, small shifts relative to ideal Bragg positions. These features arise from a variety of physical effects, including finite crystallite size, microstrain, lattice disorder and limited coherence length \cite{Warren}. Having this in mind, we develop a diffraction formalism that extends the classical Bragg picture by explicitly incorporating contributions from multiple lattice planes with non-uniform weighting. This picture creates a scattering amplitude expressed as a convergent series over harmonic components, which can be written in closed form as a polylogarithmic function. This representation introduces parameters controlling the decay of long-range correlations and the effective coherence length, allowing for a direct connection between microscopic structure and diffraction line shape. Recently, a special polylogarithm, the dilogarithm, has been successfully applied to describe the energy dependence of absorption in metal oxide films, such as V$_2$O$_5$ \cite{Saleh}, suggesting that polylogs may provide a robust framework for complex dispersion. Also the appearance of dilogarithm in diffraction problems is well-documented in the evaluation of the Maliuzhinets function \cite{Koh} and in high-energy elastic scattering models \cite{Afanasev} where the imaginary absorptive part in scattering amplitudes often expressed through logarithmic and dilogarithmic expansions.

We have further observed a key feature of this formulation. It is notable that the classical Bragg condition arises as a limiting case associated with the complex-plane structure of the polylogarithm. In particular, Bragg peaks correspond to the approach of the argument to the unit circle, where the function develops non-analytic behaviour at its branch point. Intriguingly, this perspective provides a unified interpretation of diffraction phenomena in terms of complex analysis and establishes an analogy with critical phenomena in statistical physics, where singularities arise at the boundary of analyticity.

The proposed approach does not aim to replace the Bragg law for ideal crystals, but rather to generalize it in a manner that captures the effects of finite coherence and correlation decay. As such, it could be especially relevant for the analysis of nanocrystalline, disordered or partially coherent systems, where conventional diffraction models require additional assumptions. By expressing the diffraction amplitude in terms of a polylogarithmic generating function, we provide a compact and physically transparent description of both peak positions and line shapes within a single formalism. The present approach differs from conventional peak-broadening models. This comes from the fact that the diffraction profile is not imposed phenomenologically through Gaussian or Lorentzian fitting functions, but instead emerges directly from an analytically tractable generating function. 

In Section $2$ we discuss a polylogarithm diffraction model with one brief application and the limit where the Bragg condition corresponds to the branch-cut of polylog. In Section $3$ we present an analogy of our model with critical statistical models and hyperbolic geometry. In Section $4$ we discuss about the physical relevance of our phenomenological model and its limitations. We summarise and offer a few ideas for future work in Section $5$ and \ref{app1} gives a useful calculation at Bragg limit.

\section{Polylogarithmic Diffraction from a Correlated Lattice Model}

We consider a one-dimensional lattice with sites $x_n = nd$ and density
\begin{equation}
\rho(x) = \sum_{n\in\mathbb{Z}} \eta_n \delta(x - nd)
\end{equation}
where $\eta_n$ encode structural correlations. The scattering amplitude is
\begin{equation}
F(q) = \sum_n \eta_n e^{iqnd}
\end{equation}
Introducing the pair correlation function $C(r)=\langle \eta_0 \eta_r\rangle$, the amplitude can be written as
\begin{equation}
F(q) = \sum_{r=1}^{\infty} C(r) e^{iqdr}
\end{equation}
Decomposing into real and imaginary parts,
\begin{equation}
F(q) = \sum_{r=1}^{\infty} C(r)\cos(qdr)+
i \sum_{r=1}^{\infty} C(r)\sin(qdr)
\end{equation}
An effective microscopic interpretation of this formulation may be obtained from a correlated lattice Hamiltonian \footnote{Because there is a minus sign in front of the expression, the system wants to maximize the product $\eta_n \eta_{n+r}$ in order to keep its energy as low as possible. This is closely relates to Boltzmann weight $e^{-\beta H}$. When it becomes positive, states with positive alignment are favored and for that reason we take sharp Bragg peaks.} with exponentially screened algebraic interactions,
\begin{equation}
H=-\sum_{n,r}
\frac{J_0}{r^m}
e^{-r/k_0}
\eta_n \eta_{n+r}
\label{hamiltonian}
\end{equation}
where $J_0$ is an effective coupling constant, $k_0$ plays the role of an effective coherence or correlation length measured in units of lattice spacing, while the exponent $m$ controls the algebraic suppression of long-range interference channels. Physically, finite $k_0$ may arise from crystallite boundaries, material disorders or thermal fluctuations. The limiting case $k_0\rightarrow\infty$ corresponds to restoration of long-range coherent order and ideal Bragg diffraction. Such Hamiltonians are commonly used as effective descriptions of correlated systems with finite coherence and nonlocal interactions.
Within thermal or disorder averaging, the corresponding pair correlation function takes the asymptotic form for non-ideal lattices \footnote{While exponential correlation decay is standard in diffraction theory, the additional algebraic weighting introduced here should be viewed as an effective phenomenological extension intended to capture multiscale or long-range correlated interference effects beyond conventional finite-coherence models.} which is
\begin{equation}
C(r)=\langle \eta_0 \eta_r\rangle
\sim
\frac{1}{r^m}e^{-r/k_0}
\label{corr}
\end{equation}
This form, which we use as an effective correlation kernel governing the diffraction amplitude, is mathematically a generalized Ornstein–Zernike-type decay \cite{Evans} with algebraic correction. Obviously, we have
\begin{equation}
C(r)=
\underbrace{e^{-r/k_0}}_{\text{standard}}
\;
\times
\;
\underbrace{\frac{1}{r^m}}_{\text{phenomenological}}
\end{equation}
Then the scattering amplitude becomes
\begin{equation}
F(q) = \sum_{r=1}^{\infty} \frac{e^{-r/k_0}}{r^m} e^{iqdr}
\end{equation}
Introducing $z = e^{iqd - 1/k_0}$, this becomes
\begin{equation}
F(q) = \mathrm{Li}_m(z)
\end{equation}
Hence,
\begin{align}
\Re F(q)= \sum_{r=1}^{\infty} \frac{e^{-r/k_0}}{r^m}\cos(qdr),  \\
\Im F(q) = \sum_{r=1}^{\infty} \frac{e^{-r/k_0}}{r^m}\sin(qdr)
\label{decomposition}
\end{align}
The measurable intensity is \cite{Zachariasen}
\begin{equation}
I(q) = |F(q)|^2
\end{equation}
while the imaginary part encodes phase-sensitive contributions \footnote{In experimental physics like X-Ray diffraction from multilayers \cite{Tonnerre} or crystals with anisotropic optical properties \cite{Lovesey} we have real and imaginary parts of scattering amplitude. The imaginary parts may shift the positions of Bragg peaks and the intensity of the beam.}. Within the present formulation, the Bragg condition corresponds to the approach of the polylogarithm argument toward its boundary point $z=1$, where coherent interference becomes maximized. So:
\begin{equation}
z \to 1 \quad \Longleftrightarrow \quad qd = 2\pi n
\label{condition}
\end{equation}
Near the Bragg condition, one may write (if we define a small deviation from Bragg condition $\delta=qd-2\pi n$ and expand around small quantities of $\frac{1}{k_0}$ and $\delta$ like $e^{-\frac{1}{k_0}}\approx 1-\frac{1}{k_0}$ and $e^{i\delta}=1+i\delta$)
\begin{equation}
z = e^{iqd-1/k_0}
\approx
1-\left(\frac{1}{k_0}-i(qd-2\pi n)\right)
\label{zeta}
\end{equation}
At $k_0=\infty \rightarrow \frac{1}{k_0}=0$, (\ref{zeta}) gives $z=1+i(qd-2\pi n)$ and exactly at Bragg condition (\ref{condition}) the argument approaches $z=1$ of the polylogarithm from within the unit disk. 
Consequently, the finite correlation length regularizes the singular Bragg enhancement and introduces a characteristic peak width scaling approximately as
\begin{equation}
\Delta q \sim \frac{1}{d k_0}
\end{equation}
Using $q = \frac{4\pi}{\lambda}\sin\theta$, this gives (\ref{original}).
Thus, the full scattering amplitude is governed by a polylogarithm, whose real part determines the symmetric diffraction intensity and whose imaginary part captures complex-phase (Clausen-type) contributions and also finite $k_0$ and $m$ control the deviation from ideal Bragg peaks through the singularity structure of $\mathrm{Li}_m(z)$ (\ref{app1}). Polylogarithmic structures have also appeared in other lattice-scattering systems, including coupled dipole arrays and plasmonic Fano resonances, where infinite lattice summations admit compact polylogarithmic representations \cite{Norman}.

\subsection{Determination of the Polylogarithmic Bragg Peak Shift}

\begin{figure}[h]
               \begin{center}
                \includegraphics[scale=0.34]{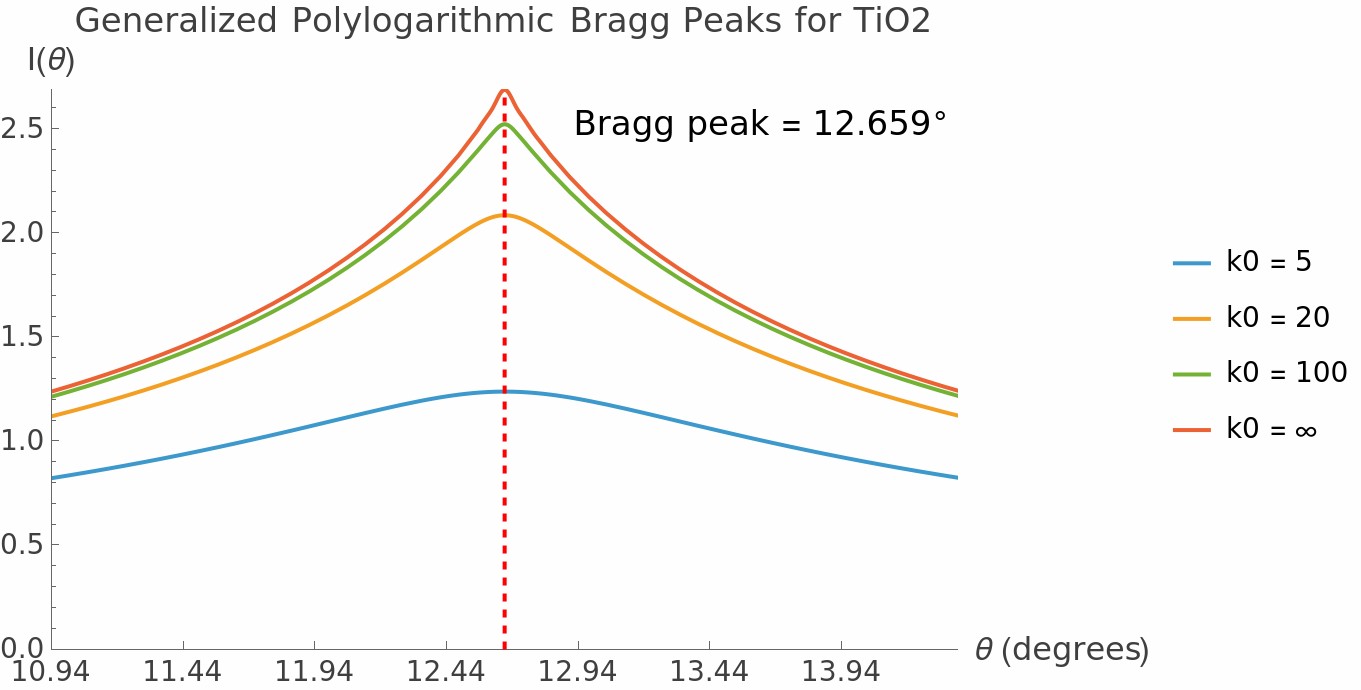}
                \end{center}
                 \caption{Polylogarithmic Bragg peak for TiO$_2$ for various values of $k_0$ parameter. We observe that finite reduced long-range order produces negligible peak-position corrections but modifies the diffraction line shape and broadening.}
                 \label{TiO2} 
                 \end{figure}

Before evaluating the numerical consequences of the present formulation, we emphasize that the main purpose of this work is to establish a theoretical and phenomenological model. It is therefore very useful to bridge this work with real-world crystallography and provide an experimental spinoff illustrative of our model, by applying our polylogarithmic framework as a localized case study along a specific high-symmetry direction of a well-characterized material.
For anatase TiO$_2$ \footnote{In the realistic 3D case of anatase, the scattering amplitude generalizes to a sum over the 3D lattice vectors $\mathbf{R}_{\mathbf{n}}$:
\begin{equation}
F(\mathbf{q}) = \sum_{\mathbf{n}} \eta(\mathbf{R}_{\mathbf{n}}) e^{i \mathbf{q} \cdot \mathbf{R}_{\mathbf{n}}}
\end{equation}
where $\mathbf{q}$ is the scattering vector. For a specific $(hkl)$ reflection, the 3D correlation decay $C(\mathbf{R}) \sim e^{-R/k_0} R^{-m}$ reveals a polylogarithmic representation along the direction of the reciprocal lattice vector $\mathbf{G}_{hkl}$. }, a technologically important photocatalytic oxide material with tetragonal crystal structure, with lattice spacing $d_{101}=3.515$\,\AA\ and $(\text{Cu K}\alpha)$ radiation $\lambda=1.5406$\,\AA, the standard Bragg condition (\ref{original})
gives the first-order diffraction angle
$
\theta_B \approx 12.66^\circ.
$
Experimentally, X-ray diffraction patterns are conventionally plotted as a function of $2\theta$, so the physical Bragg peak appears at
$
2\theta_B \approx 25.32^\circ,
$
which is the well-known anatase $(101)$ diffraction peak.
In the polylogarithmic model for $\theta_{\mathrm{eff}}=qd$ we have
\begin{equation}
F(\theta)=\sum_{r=1}^{\infty}\frac{e^{-r/k_0}}{r^m}e^{ir\theta_{\mathrm{eff}}}
\end{equation}
the parameters $k_0$ and $m$ do not significantly modify the Bragg angle itself, but instead affect line shape and peak broadening. Consequently in this case, the physically meaningful prediction of the proposed formulation is not a shift of the Bragg position, but a polylogarithmic description of the diffraction line shape and coherence structure in correlated or disordered lattices. The polylogarithm Bragg peaks for TiO$_2$ are on (\ref{TiO2}), for $m=2$ and for various values of $k_0$.

Nevertheless, the present formulation should be viewed as complementary to conventional diffraction broadening approaches such as the Debye--Waller factor, Scherrer analysis, or Williamson--Hall methods \cite{Rene}. While these techniques successfully parameterize peak broadening phenomenologically, the polylogarithmic framework derives the diffraction profile directly from an underlying correlation structure, allowing coherence length and harmonic suppression to emerge naturally from the analytic form of the scattering amplitude.

We define the intensity as
\begin{equation}
I(\theta) = |F(\theta)|^2
=
\left|
\sum_{r=1}^{\infty}
\frac{e^{-r/k_0}}{r^m}
e^{ir\alpha \sin\theta}
\right|^2,
\qquad
\alpha = \frac{4\pi d}{\lambda}.
\end{equation}
For TiO$_2$, $\alpha \approx 28.67$ and the Bragg peak is obtained numerically via discretization of the intensity function. The infinite sum is truncated at $r=200$:
\begin{equation}
F(\theta) \approx \sum_{r=1}^{200}
\frac{e^{-r/k_0}}{r^m}
e^{ir\alpha\sin\theta}
\end{equation}
For illustrative numerical evaluation, we use the parameters:
$m = 2,
k_0 = 50.${\footnote{
The values $m=2$ and $k_0=50$ are chosen for illustrative and numerical stability purposes. The exponent $m=2$ leads to a dilogarithmic structure factor $\mathrm{Li}_2(z)$, which possesses the well-known analytical and geometric properties and coherence scale $k_0=50$ represents a moderately long but finite correlation length, sufficiently large to reproduce sharp Bragg-like diffraction while still allowing observable finite-coherence broadening effects. These values are not intended to represent universal material constants and should be extracted phenomenologically through fitting to experimental diffraction profiles.
}}
The intensity is evaluated on a dense angular grid:
\begin{equation}
\theta \in [\theta_B - \Delta, \theta_B + \Delta],
\qquad \Delta \sim 0.1\,\text{rad}
\end{equation}
The polylog Bragg peak is determined as
$
\theta_{\text{peak}} = \arg\max_{\theta_j} I(\theta_j),
$
yielding the numerical result
$
\theta_{\text{peak}} = 12.6618^\circ.
$
The numerical maximum differs only slightly from the classical Bragg angle,
\begin{equation}
\delta\theta = \theta_{\text{peak}} - \theta_B
= 0.00287^\circ
\end{equation}
indicating that finite coherence damping primarily affects the peak width and line shape rather than the Bragg position itself.

\subsection{Decomposition of $F(q)$ and the imaginary component}
We have shown in (\ref{decomposition}) the decomposition of $F(q)$ amplitude in two parts 
$\Re F(q)$ and $\Im F(q)$
follows directly from
\begin{equation}
F(q)=\mathrm{Li}_m\!\left(e^{iqd-1/k_0}\right)
\end{equation}
The measurable diffraction intensity is determined by
\begin{equation}
I(q)=|F(q)|^2
=(\Re F(q))^2+(\Im F(q))^2
\end{equation}
For centrosymmetric crystals such as anatase TiO$_2$, the diffraction pattern is predominantly governed by the real part of the scattering amplitude. In this case,
$
\Re F(q)
$
describes the coherent symmetric interference responsible for the observed Bragg peak structure. Near the Bragg condition
$
qd=2\pi n,
$
the cosine terms interfere constructively,
\begin{equation}
\cos(2\pi nr)=1
\end{equation}
leading to a strong enhancement of the diffraction intensity.

By contrast, the imaginary part
$
\Im F(q)
$
encodes the antisymmetric component of the scattering amplitude. For an ideal centrosymmetric lattice, this contribution is expected to vanish or remain strongly suppressed near the principal Bragg peak because
\begin{equation}
\sin(2\pi nr)=0
\end{equation}
Consequently, for the TiO$_2$ $(101)$ reflection one expects:
\begin{align}
\Re F(q) &\rightarrow \text{dominant sharp maximum near } 2\theta_B \approx 25.3^\circ, \nonumber \\
\Im F(q) &\rightarrow \text{nearly vanishing contribution near the peak}.
\end{align}
On the other hand it is extremely interesting that for Bragg crystals with arbitrary thickness, dynamical Bragg reflection and transmission induced only by the imaginary part of the atomic scattering factor \cite{Zhangcheng}. Finite coherence suppresses higher-order interference contributions and broadens the diffraction profile, while the exponent $m$ controls the relative weight of higher-order interference channels. Thus, the polylogarithmic structure factor provides a unified analytic description of coherent diffraction, peak broadening, and correlation effects in non-ideal lattices.

Since the imaginary component (\ref{decomposition}) is typically suppressed in ordinary centrosymmetric crystals because the diffraction process is dominated by symmetric correlations, for conventional materials such as anatase TiO$_2$, the diffraction intensity is governed primarily by the real part of the polylogarithmic structure factor.
However, the imaginary sector may become physically relevant in systems with broken inversion symmetry, chiral order, or complex phase-modulated correlations. In such systems, the scattering amplitude acquires a nontrivial phase structure and the contribution
\begin{equation}
\Im\left[\mathrm{Li}_m(e^{iqd-1/k_0})\right]
\end{equation}
can influence the diffraction response. This phase-sensitive sector has conceptual similarity to diffuse scattering contributions observed in quasicrystalline and aperiodic systems, where nontrivial correlation phases and finite coherence effects modify the ideal Bragg response. Geometrically, this correspondence motivates a formal analogy between phason-induced diffuse scattering and the Bloch--Wigner dilogarithm, whose imaginary sector is known to encode hyperbolic volumes. In this interpretation, the diffuse contribution may formally be associated with a phase-sensitive sector of the generalized polylog.

Noticing that we begin to suspect that examples that could possibly fit in our polylogarithm structure of Bragg diffraction could include quasicrystals and aperiodic lattices, incommensurate charge-density-wave systems, chiral crystals, twisted or moir\'e superlattices, magnetic helices, spiral spin systems and topological photonic or phononic lattices. In these cases, the imaginary part encodes phase-sensitive interference effects and modulation-induced diffraction structures that are absent in ordinary periodic crystals. The polylogarithmic interpretation therefore becomes most relevant not for ideal Bragg diffraction, but for correlated and phase-structured lattices where complex interference effects play a central role.

 \section{Analogy with critical phenomena in statistical physics and connection to hyperbolic geometry}
 The polylogarithmic diffraction amplitude $F(\theta) = \mathrm{Li}_m(z)$, exhibits a direct analogy with critical phenomena in statistical physics, so it is therefore useful to identify possible universal patterns. In both cases, the polylogarithm develops non-analytic behaviour at the $z = 1$, which corresponds to the critical point. In the diffraction context, this condition is equivalent to $qd=\theta_{\mathrm{eff}} = 2\pi n$, i.e., the Bragg condition. The parameter $\epsilon = 1/k_0$ plays the role of an inverse coherence length, analogous to the inverse correlation length in critical systems. As $k_0 \to \infty$, the system approaches the unit circle in the complex plane, and the diffraction amplitude develops singular behaviour, producing sharp Bragg peaks. Near this point, the amplitude $F(\theta)$ is a function of $\epsilon -i(\theta - \theta_B)$, implying that the peak width scales as $\Delta\theta \sim 1/k_0$. Thus, Bragg diffraction exhibits a formal analogy with critical phenomena with polylogarithm basis.

Having this idea in mind, an interesting analogy follows between the present diffraction formalism and finite-temperature quantum field theory, in particular the Gross--Neveu model at imaginary chemical potential \cite{Filothodoros2016, Filothodoros2018}. In both cases, the relevant quantities can be expressed in terms of polylogarithmic functions with arguments of the form $z = e^{-A + iB}$. In the diffraction context, one has $z = e^{i\theta_{\mathrm{eff}} - \epsilon}$, with $\theta_{\mathrm{eff}}$ representing an angular variable on the boundary and $\epsilon$ controlling the radial distance from the boundary. In the Gross--Neveu model, the analogous parameter is $z = e^{-\beta(\sigma - i\theta)}$, where $i\theta$ is the imaginary chemical potential and $\sigma$ is the dynamically generated mass. The real and imaginary parts of the polylogarithm play distinct roles in both settings: the real part governs symmetric observables such as diffraction intensity or thermodynamic free energy, while the imaginary part, related to Clausen functions, captures phase-sensitive or geometric contributions. In both systems, non-analytic behaviour arises as $z$ approaches the unit circle, corresponding respectively to the emergence of Bragg peaks and to critical or conformal behavior in the field theory. This parallel admits a deeper unifying role of polylogarithmic analytic structure across seemingly disparate physical domains where remarkably Bragg peaks and critical thermodynamic behaviour arise from the same underlying mechanism: the boundary behaviour of polylogarithmic function. Interestingly, in \cite{Cheng} the identification of Bragg peaks with analogous branch points of polylogarithmic functions is supported in subwavelength grating diffraction. So the overall picture is

\begin{table}[h]
\centering
\resizebox{\columnwidth}{!}{%
\begin{tabular}{|c|c|c|}
\hline
\textbf{Sector} & \textbf{Diffraction} & \textbf{Gross--Neveu} \\
\hline
$\Re\big[\mathrm{Li}_m(z)\big]$ & Intensity & Thermodynamics \\
\hline
$\Im\big[\mathrm{Li}_m(z)\big]$ & Phase-sensitive & Geometric \\
\hline
\end{tabular}%
}
\caption{Parallel interpretation of the polylogarithm sectors in diffraction theory and the Gross--Neveu model at imaginary chemical potential.}
\label{tab:polylog_correspondence}
\end{table}
Also, a deeper geometric interpretation of the generalized diffraction formalism appears in the special case $m=2$, where the scattering amplitude is governed by the dilogarithm,
\begin{equation}
F(\theta) = \mathrm{Li}_2(z), \qquad z = e^{i\theta_{\mathrm{eff}} - \epsilon}
\end{equation}
where $\theta_{\mathrm{eff}}$ and $\epsilon$ are angular and radial parameters as we have mentioned before. This construction is formally analogous to the geometric structure of the Poincar\'e disk representation of hyperbolic space, where the unit circle plays the role of a conformal boundary and the interior of the disk is the bulk.
In this context, we introduce the Bloch--Wigner dilogarithm \cite{Zagier1, Zagier2},
\begin{equation}
D(z) = \Im\left[\mathrm{Li}_2(z)\right] + \arg(1 - z)\log|z|
\end{equation}
which is a real-valued and single-valued function with a well-known geometric interpretation: it gives the volume of an ideal tetrahedron in hyperbolic three-space. It is interesting that in \cite{Cherqui}, for $1D$ nanoparticle arrays when lattice sums are taken over an infinite number of interactions they yield closed-form analytic expressions involving polylogs. Crucially, the emergence of the sharp Bragg mode in these systems is associated with the limiting behaviour of these polylogarithmic features, providing a physical foundation for treating Bragg diffraction as a boundary phenomenon arising from the boundary behaviour of an analytic generating function.

The complex parameter $z$ can be viewed as a coordinate in the unit disk. In the limit $\epsilon \to 0$, one has $z \to e^{i\theta_{\mathrm{eff}}}$, corresponding to an approach to the boundary of hyperbolic space. In this limit, the imaginary part of the dilogarithm reduces to the Clausen function \cite{Lewin},
\begin{equation}
\Im\left[\mathrm{Li}_2(e^{i\theta})\right] = Cl_2(\theta)
\end{equation}
which therefore admits a direct interpretation in terms of hyperbolic volume.

The $z \to 1$ point corresponds simultaneously to a singular limit of the polylogarithmic generating function and to a degenerate configuration in hyperbolic geometry. Thus, diffraction peaks may be interpreted in terms of geometric singularities associated with the analytic structure of the Bloch--Wigner dilogarithm, establishing a connection between lattice interference, special functions, and hyperbolic geometry.

\begin{equation}
\resizebox{1.02\columnwidth}{!}{%
$\boxed{\text{Bragg Peak} \longleftrightarrow \text{Boundary Singularity in Hyperbolic Geometry}}$%
}
\end{equation}
So while quasicrystals and conventional lattices are typically described geometrically, systems with power-law and finite-range correlations give rise naturally to polylogarithmic structure factors, suggesting an alternative classification of diffraction phenomena based on analytic generating functions rather than spatial order alone.

\section{Physical Relevance and Limitations}
The generalized diffraction formula based on the polylogarithmic generating function is most useful in contexts where deviations from ideal crystalline order play a significant role. In particular, it provides a physically motivated framework for analyzing diffraction line shapes in nanocrystalline materials, disordered lattices, and systems with finite coherence length, where peak broadening, asymmetry, and subtle shifts cannot be fully captured by the standard Bragg law. By encoding correlation decay through parameters such as $k_0$ and $m$, the model enables a direct connection between diffraction data and underlying structural correlations.
In such a setup, these parameters should be interpreted as emergent coherence characteristics extracted from diffraction behaviour. 

However, in the limit of perfect periodicity and long-range order of ideal crystals, where coherence is effectively infinite, the formulation reduces to the conventional Bragg condition and does not offer a practical advantage. Consequently, for routine crystallographic applications such as phase identification or lattice parameter determination, the polylog expression is unnecessary and standard Bragg-based methods remain sufficient.
We may claim that the present model remains phenomenological in the sense that the specific correlation kernel is introduced as an effective description rather than derived from a fully microscopic many-body treatment of a particular material. Nevertheless, the generalized correlation form may be motivated by effective lattice Hamiltonians with exponentially screened algebraic interactions, which naturally generate finite-coherence and long-range correlation effects. So, the present work should be viewed primarily as an analytically tractable theoretical framework intended to explore how generalized correlation structures modify conventional diffraction behaviour. 

\section{Conclusion}
We have developed a polylog diffraction framework that extends the classical Bragg description by incorporating correlation-dependent weighting of multi-plane interference contributions. Using this weighting, we extend the Bragg condition to include the effects of finite coherence directly within the scattering amplitude. This is made possible through a polylogarithmic generating function.

Within this approach, which was led by mathematical curiosity, the classical Bragg condition is recovered in the long-coherence limit, while finite values of the coherence scale $k_0$ produce diffraction features associated with partial suppression of long-range interference. The proposed structure factor therefore provides a unified analytic framework connecting ideal Bragg diffraction with finite-coherence scattering in correlated or disordered lattices.

More generally, the present formulation suggests a possible connection between generalized diffraction theory and long-range statistical lattice models. Correlation kernels of the form
\begin{equation}
C(r)\sim \frac{e^{-r/k_0}}{r^m}
\end{equation}
naturally appear in long-range Ising- and XY-type systems, where the parameters $m$ and $k_0$ control the suppression of distant correlations and the effective coherence scale. Within this perspective, the polylogarithmic diffraction amplitude may be interpreted as the structure-factor analogue of a long-range correlated lattice theory, while the singular behaviour near $z\rightarrow1$ formally resembles the nonanalytic structure associated with critical phenomena and phase-coherent ordering.

In addition, the appearance of polylogarithmic and Clausen-type functions indicates possible connections between diffraction theory, critical phenomena, and geometric structures.
We expect that future extensions of the present work may include anisotropic correlation functions, higher-dimensional lattices, and direct comparison with diffraction data from quasicrystals, chiral magnetic systems, and moir\'e superlattices.

\subsection*{Acknowledgements} I would like to thank Anastasios Petkou for helpful discussion.

\appendix
\section{Asymptotic Structure Near the Bragg Limit}
\label{app1}

The polylogarithmic representation allows a direct analytical study of the diffraction amplitude close to the Bragg condition. We define
\begin{equation}
z=e^{iqd-1/k_0}
=
e^{-(\epsilon-i\delta)}
\end{equation}
with
\begin{equation}
\epsilon=\frac{1}{k_0},
\qquad
\delta=qd-2\pi n
\end{equation}
where $\delta$ measures the deviation from the $n$th Bragg condition. The Bragg limit corresponds to
\begin{equation}
\epsilon \to 0,
\qquad
\delta \to 0,
\qquad
z\to1
\end{equation}
The diffraction amplitude becomes
\begin{equation}
F(q)=\mathrm{Li}_m(e^{-(\epsilon-i\delta)})
\end{equation}
For small complex argument
\begin{equation}
\mu=\epsilon-i\delta
\end{equation}
the polylogarithm admits the asymptotic expansion \cite{Lewin}
\begin{equation}
\mathrm{Li}_m(e^{-\mu})
=
\Gamma(1-m)\mu^{m-1}
+
\sum_{k=0}^{\infty}
\frac{(-\mu)^k}{k!}\zeta(m-k)
\label{polyasym}
\end{equation}
valid for $|\mu|<2\pi$ through analytic continuation. For integer $m>1$, logarithmic corrections appear in the analytic continuation of (\ref{polyasym}).

Substituting the complex variable $\mu = \epsilon - i\delta$ into the asymptotic expansion of the dilogarithm ($m=2$), we can evaluate the behaviour of the scattering amplitude as the system approaches the boundary of perfect coherence. In the limit $\epsilon \to 0$, the expansion of $F(q)$ around the point $z=1$ becomes:

\begin{equation}
F(q) \approx \frac{\pi^2}{6} + i\delta \left( \ln(i\delta) - 1 \right)
\end{equation}
Using the identity $\ln(i\delta) = \ln|\delta| + i\frac{\pi}{2} \operatorname{sgn}(\delta)$, we decompose the amplitude into its real and imaginary parts:
\begin{equation}
\Re[F(q)] \approx \frac{\pi^2}{6} -|\delta|\frac{\pi}{2}, \quad \Im[F(q)] \approx \delta \ln|\delta|-\delta
\label{asymptotic_result}
\end{equation}
This result indicates that the real part $\Re[F(q)]$ exhibits a cusp-like maximum at the Bragg condition ($\delta=0$) and the imaginary part $\Im[F(q)]$ recovers the asymptotic form of the second-order Clausen function $\text{Cl}_2(\delta) \sim \delta \ln|\delta|-\delta$. The logarithmic nonanalyticity at $z \rightarrow 1$ makes the phase sector more sensitive to small deviations from the Bragg condition, directly connecting the branch-cut geometry to the diffusion signs observed in partially ordered or quasi-crystals.


\begin{thebibliography}{00}

\bibitem{Egami}
Takeshi Egami, and Simon J.L. Billinge,
\newblock Underneath the Bragg Peaks, Structural Analysis of Complex Materials,
\newblock \textit{Pergamon Materials Series} (2003)

\bibitem{Warren}
B. E. Warren,
\newblock X-ray Diffraction,
\newblock \textit{Dover Publications} (1990)

\bibitem{Saleh}
Saleh, M.H., Al-Daraghmeh, T.M., Bulos, B. et al.,
\newblock Validity of a three-parameter dilogarithm dispersion function for the description of the energy dependence of the absorption coefficient of flash-evaporated V$_2$0$_5$,
\newblock \textit{Discov Mater} \textbf{6} (2026)

\bibitem{Koh}
Il-Suek Koh, and Yongshik Lee,
\newblock Exact Evaluation of the Maliuzhinets Half-Plane Function,
\newblock \textit{ IEEE Transactions on Antennas and Propagation} \textbf{55} (2007)

\bibitem{Afanasev}
Andrei V. Afanasev and N. P. Merenkov,
\newblock Large logarithms in the beam normal spin asymmetry of elastic electron-proton scattering,
\newblock \textit{Phys. Rev. D} \textbf{70} (2004)

\bibitem{Evans}
R Evans and J R Henderson,
\newblock Pair correlation function decay in models of simple fluids that contain dispersion interactions,
\newblock \textit{Journal of Physics: Condensed Matter} \textbf{21} (2009)


\bibitem{Zachariasen}
Zachariasen, William H.,
\newblock Theory Of X-ray Diffraction In Crystals,
\newblock \textit{J. Wiley and sons, inc., Chapman and Hall, ltd.} (1945)

\bibitem{Norman}
Justin C. Norman, Drew F. DeJarnette, and D. Keith Roper,
\newblock Polylogarithm-Based Computation of Fano Resonance in Arrayed
Dipole Scatterers,
\newblock \textit{The Journal of Physical Chemistry C} \textbf{ACS Publications} (2014)

\bibitem{Tonnerre}
L. Sève, J. M. Tonnerre, D. Raoux,
\newblock Determination of the Anomalous Scattering Factors in the Soft-X-ray Range using Diffraction from a Multilayer,
\newblock \textit{J. Appl. Cryst.} \textbf{31} (1998)

\bibitem{Lovesey}
S W Lovesey, V Scagnoli, A N Dobrynin, Y Joly and S P Collins,
\newblock Effects of dispersion and absorption in resonant Bragg diffraction of x-rays,
\newblock \textit{Journal of Physics: Condensed Matter} \textbf{26} (2014)

\bibitem{Rene}
René Guinebretière,
\newblock X‐ray Diffraction by Polycrystalline Materials,
\newblock \textit{Wiley} (2007)

\bibitem{Zhangcheng}
Xu Zhangcheng et al,
\newblock Bragg reflection and transmission of X-rays induced by the imaginary part of the atomic scattering factor, 
\newblock \textit{ J. Phys.: Condens. Matter} \textbf{7} (1995)


\bibitem{Filothodoros2016}
E.~G. Filothodoros, A.~C. Petkou, and N.~D. Vlachos,
\newblock $3d$ fermion-boson map with imaginary chemical potential,
\newblock \textit{Phys. Rev. D} \textbf{95} (2017), no.~6 065029


\bibitem{Filothodoros2018}
Evangelos G. Filothodoros, Anastasios C. Petkou, Nicholas D. Vlachos,
\newblock The fermion–boson map for large d,
\newblock \textit{Nuclear Physics B} \textbf{941} (2019)

\bibitem{Cheng}
Y. C. Cheng, J. Redondo, and K. Staliunas,
\newblock Beam focusing in reflections from flat subwavelength diffraction gratings,
\newblock \textit{Phys. Rev. A} \textbf{89} (2014)

\bibitem{Zagier1}
D.~Zagier,
\newblock The Bloch-Wigner-Ramakrishnan polylogarithm function,
\newblock \textit{Math. Ann.} \textbf{286} (1990), no.~1-3 613–624


\bibitem{Zagier2}
D.~Zagier,
\newblock The dilogarithm function,
\newblock \textit{Frontiers in Number Theory, Physics and Geometry II},
P. Cartier, B. Julia, P. Moussa, P. Vanhove (eds.),
Springer-Verlag, Berlin-Heidelberg-New York (2006) 3–65

\bibitem{Cherqui}
Charles Cherqui, Marc R.Bourgeois, Danqing Wang, and George C. Schatz,
\newblock Plasmonic Surface Lattice Resonances:Theory and Computation,
\newblock \textit{Acc. Chem. Res.} \textbf{52} (2019)


\bibitem{Lewin}
Leonard Lewin,
\newblock Polylogarithms and Associated Functions,
\newblock \textit{North Holland} (1981)

\end{thebibliography}
\end{document}